%% file: coalition-sr.tex
\documentclass[submission,copyright]{eptcs}
\providecommand{\event}{SR 2014} 
\usepackage{breakurl}             
\usepackage{amssymb,amsmath}
\usepackage{setspace}
\usepackage{stmaryrd} 
\usepackage{xcolor}
\usepackage{enumitem}
\usepackage{xspace}
\usepackage{graphicx}

\SetSymbolFont{stmry}{bold}{U}{stmry}{m}{n}

\newtheorem{lemma}{Lemma}
\newtheorem{theorem}{Theorem}
\newcommand{\clprover}{{\sf\bf CLProver}\xspace}

\input{symbols.tex}

\title{A Resolution Prover for Coalition Logic}

\author{Cl\'audia Nalon
\institute{Departament of Computer Science\\
           University of Bras\'{\i}lia (Brazil)\\
\email{nalon@unb.br}}
\and
Lan Zhang
\institute{Information School\\
           Capital University of Economics and Business (China)\\
           \email{lan@cueb.edu.cn}}
\and
Clare Dixon\quad\quad Ullrich Hustadt
\institute{ Department of Computer Science\\
            University of Liverpool (UK) \\
          \email{$\{$CLDixon,U.Hustadt$\}$@liverpool.ac.uk}}
}
\def\titlerunning{A Resolution Prover for \protect\system{CL}}
\def\authorrunning{C. Nalon, L. Zhang, C. Dixon \& U. Hustadt}
\begin{document}
\maketitle

\begin{abstract}
We present a prototype tool for automated reasoning for Coalition Logic,
a non-normal modal logic that can be used for reasoning about
cooperative agency. The theorem prover \clprover is based on recent work
on a resolution-based calculus for Coalition Logic that operates on
coalition problems, a normal form for Coalition Logic. We provide an
overview of coalition problems and of the resolution-based calculus for
Coalition Logic. We then give details of the implementation of \clprover and present the results for a comparison with an
existing tableau-based solver.
\end{abstract}

\section{Introduction}

Coalition Logic \system{CL}{}{} is a formalism intended to describe the ability of groups of agents to achieve an outcome in a strategic game \cite{Pauly2002}. 
\system{CL}{}{} is a multi-modal logic with modal operators of the form \coop{A}, where $\set{A}$ is a set of agents. The formula $\coop{A}\varphi$ reads as \emph{the coalition $\set{A}$ has a strategy to achieve $\varphi$}, where $\varphi$ is a formula. We note that \system{CL}{}{} is a non-normal modal logic, as the schema that represents \emph{additivity}, $\coop{A}\varphi \land \coop{A}\psi \then \coop{A}(\varphi \land \psi)$, is not valid. However, \emph{monotonicity}, $\coop{A}(\varphi \land \psi) \then \coop{A}\varphi \land \coop{A}\psi$, holds.

Coalition Logic is equivalent to the next-time fragment of \emph{Alternating-Time Temporal Logic} (\system{ATL}{}{}) \cite{AHK1997,Goranko2001}, where $\coop{A}\varphi$ translates into $\excoa{A}\Next\varphi$ (read as \emph{the coalition $\set{A}$ can ensure $\varphi$ at the next moment in time}). The satisfiability problems for \system{ATL}{}{} and \system{CL}{}{} are EXPTIME-complete \cite{WLWW2006} and PSPACE-complete \cite{Pauly2002}, respectively. Proof methods for these logics include, for instance, tableau-based methods for \system{ATL}{}{} \cite{WLWW2006,GS2009} and a tableau-based method for \system{CL}{}{} \cite{Hansen2004}. 

In order to make the paper self-contained, we present here the resolution-based calculus for \system{CL}{}{}, \calculus{CL} \cite{NZDH-JLC-2014}. As to the best of our knowledge, there are no other resolution-based methods for either \system{ATL}{}{} or \system{CL}{}{}. Providing such a method for \system{CL}{}{} gives the user a choice of proof methods. Several comparisons of tableau algorithms and resolution methods \cite{HustadtSchmidt02b,GTW2011} indicate that there is no overall best approach: for some classes of formulae tableau algorithms perform better whilst on others resolution performs better. So, with a choice of different provers, for the best result, the user could run several in parallel or the one most likely to succeed depending on the type of the input formulae. \calculus{CL} is sound, complete, and terminating as shown in \cite{NZDH-JLC-2014}.

The paper is organised as follows. In the next section, we present the syntax, axiomatisation, and semantics of \system{CL}{}{}. In Section~\ref{sec:calculus}, we introduce the resolution-based method for \system{CL}{}{}, the main results, and provide a small example. In Section~\ref{sec:implementation}, we introduce the theorem-prover for \system{CL}{}{}. We give details of the implementation and discuss the results for a comparison with an existing tool. Conclusions and future work are given in Section~\ref{sec:conclusion}. 

\section{Coalition Logic}\label{sec:language}

As in \cite{GS2009}, we define $\allag \subset \Nat$ to be a finite, non-empty set of agents. A \define{coalition} $\set{A}$ is a subset of $\allag$. Formulae in \system{CL}{}{} are constructed from propositional symbols ($\Prop = \Set{p,q,r,\ldots,p_1,q_1,r_1,\ldots}$) and constants ($\constant{true}, \constant{false}$), together with Boolean operators ($\neg$, for negation, and $\land$, for conjunction) and coalition modalities. Formulae whose main operator is classical are built in the usual way. A \define{coalition modality} is either of the form $\coop{A}\varphi$ or $\dualcoop{A}\varphi$, where $\varphi$ is a well-formed \system{CL}{}{} formula. The coalition operator $\dualcoop{A}$ is the dual of $\coop{A}$, that is, $\dualcoop{A}\varphi$ is an abbreviation for $\neg \coop{A}\neg\varphi$, for every coalition $\set{A}$ and formula $\varphi$. We denote by \WFF{CL} the set of \system{CL}{}{} well-formed formulae. Parentheses will be omitted if the reading is not ambiguous. We also omit the curly brackets within modalities. For instance, we write $\coopl{1,2}\varphi$ instead of $\coopl{\{1,2\}}\varphi$. Formulae of the form $\bigvee \varphi_i$ (resp.\ $\bigwedge \varphi_i$), $1\leq i \leq n$, $n \in \Nat$, $\varphi_i \in \WFF{CL}$, represent arbitrary disjunctions (resp.\ conjunctions) of formulae. If $n=0$, $\bigvee \varphi_i$ (resp.\ $\bigwedge \varphi_i$) is called the \define{empty disjunction} (resp.\ \define{empty conjunction}), denoted by $\constant{false}$ (resp.\ $\constant{true}$). 

A \define{literal} is either $p$ or $\neg p$, for $p \in \Prop$. For a literal $l$ of the form $\neg p$, where $p$ is a propositional symbol, $\neg l$ denotes $p$; for a literal $l$ of the form $p$, $\neg l$ denotes $\neg p$. The literals $l$ and $\neg l$ are called \define{complementary literals}. We assume that literals are in simplified form, that is, $\neg\neg l$ is assumed to be $l$. A \define{positive coalition formula} (resp.\ \define{negative coalition formula}) is a formula of the form $\coop{A}\varphi$ (resp.\ $\dualcoop{A}\varphi$), where $\varphi \in \WFF{CL}$. A \define{coalition formula} is either a positive or a negative coalition formula.

Coalition logic can be axiomatised by the following schemata (where $\set{A}, \set{A'}$ are coalitions and $\varphi, \varphi_1$, $\varphi_2$ are well-formed formulae) \cite{Pauly2002}:

\[
\begin{array}{lcl}
\axiom{\bot} &:& \neg\coop{A}\constant{false} \\
\axiom{\top} &:& \coop{A}\constant{true}\\
\axiom{\mbox{\allag}} &:& \neg \coopl{\emptyset}\neg \varphi \then \coopl{\allag}\varphi \\
\axiom{M} &:& \coop{A}(\varphi_1 \land \varphi_2) \then \coop{A}\varphi_1\\
\axiom{S} &:& \coop{A}\varphi_1 \land \coop{A'}\varphi_2 \then \coopl{\set{A}\cup\set{A}'}(\varphi_1 \land\varphi_2), \mbox{ if } \set{A}\cap\set{A}' = \emptyset
\end{array}
\]

\noindent together with propositional tautologies and the following inference rules: \rulename{modus ponens} (from $\varphi_1$ and $\varphi_1 \then \varphi_2$ infer $\varphi_2$) and \rulename{equivalence} (from $\varphi_1 \ifonlyif \varphi_2$ infer $\coop{A}\varphi_1 \ifonlyif \coop{A}\varphi_2$). It can be shown that the inference rule \rulename{monotonicity} (from $\varphi_1 \then \varphi_2$ infer $\coop{A}\varphi_1 \then \coop{A}\varphi_2$) is a derivable rule in this system. The next result will be used later.

\begin{lemma}\label{lemma:axiomatisation} The formula $\coop{A}\psi_1 \land \dualcoop{B} \psi_2\then \dualcoopl{\set{B}\setminus\set{A}}(\psi_1 \land \psi_2)$ where $\set{A}$ and $\set{B}$ are coalitions, $\set{A}\subseteq\set{B}$, and $\psi_1, \psi_2 \in \WFF{CL}$, is valid.
\end{lemma}

\begin{lproof}
\begin{tabular}{l@{\hskip 2ex}l@{\hskip 4ex}l}
1. & $\coop{A}\psi_1 \land \coopl{\set{B}\setminus\set{A}}(\psi_1 \then \neg \psi_2) \then \coop{B}(\psi_1 \land ( \psi_1 \then \neg \psi_2))$ & $\axiom{S}, \set{A}' = \set{B}\setminus\set{A}$\\
  &                                                            & $\varphi_1=\psi_1, \varphi_2 = \psi_1 \then \neg \psi_2$\\

2. & $\psi_1 \land (\psi_1 \then \neg \psi_2) \then \neg \psi_2$ & \emph{propositional tautology}\\

3. & $\coop{B}(\psi_1 \land (\psi_1 \then \neg \psi_2)) \then \coop{B}\neg \psi_2$ & 2, \emph{monotonicity} \\

4. & $\coop{A}\psi_1 \land \coopl{\set{B}\setminus\set{A}}(\psi_1 \then \neg \psi_2)\then \coop{B} \neg \psi_2$ & 1,3, \emph{chaining}\\

5. & $\coop{A}\psi_1 \land \neg \coop{B} \neg \psi_2\then \neg \coopl{\set{B}\setminus\set{A}}(\neg \psi_1 \lor \neg \psi_2)$ & 4, \emph{rewriting}\\

6. & $\coop{A}\psi_1 \land \dualcoop{B} \neg\neg \psi_2\then \dualcoopl{\set{B}\setminus\set{A}}\neg(\neg \psi_1 \lor \neg \psi_2)$ & 5, \emph{def. dual}\\

7. & $\coop{A}\psi_1 \land \dualcoop{B} \psi_2\then \dualcoopl{\set{B}\setminus\set{A}}(\psi_1 \land \psi_2)$ & 6, \emph{rewriting}\hfill$\Box$\\
\end{tabular}
\end{lproof}

\vspace{2ex}The semantics of \system{CL}{}{} is given in terms of \emph{Concurrent Game Structures} (CGS) \cite{AHK2002} and it is \emph{positional}, that is, agents have no memory of their past decisions and, thus, those decisions are made by taking into account only the current state. We note that the semantics of \system{CL}{}{} is often presented in terms of \emph{Multiplayer Game Models} (MGMs) \cite{Pauly2001}. Note also that MGMs yield the same set of validities as CGSs \cite{Goranko2001}. As we intend to extend the proof method given here to full \system{ATL}{}{}, the correctness proofs are based on the tableau procedure for full \system{ATL}{}{} \cite{GS2009} and we follow the semantics presentation given there.

\begin{definition} A \define{Concurrent Game Frame} (CGF) is a tuple $\frames{F} = \Frames{\allag,\set{S},s_0,d,\delta}$, where
\begin{itemize}[noitemsep]
\item $\allag$ is a finite non-empty set of \define{agents};
\item $\set{S}$ is a non-empty set of \define{states}, with a distinguished state $s_0$, termed \emph{initial state};
\item $d: \allag \times \set{S} \longrightarrow \Nat^+$, where the natural number $d(a,s)\geq 1$ represents the \define{number of moves} that the agent $a$ has at the state $s$. Every \define{move} for agent $a$ at the state $s$ is identified by a number between $0$ and $d(a,s)-1$. Let $D(a,s)=\Set{0,\ldots,d(a,s)-1}$ be the set of all moves available to agent $a$ at $s$. For a state $s$, a \define{move vector} is a $k$-tuple $\Tuple{\sigma_1,\ldots,\sigma_k}$, where $k=|\allag|$, such that $0 \leq \sigma_a \leq d(a,s)-1$, for all $a \in \allag$. Intuitively, $\sigma_a$ represents an arbitrary move of agent $a$ in $s$. Let $D(s) = \Pi_{a \in \allag} D(a,s)$ be the set of all move vectors at $s$. We denote by $\sigma$ an arbitrary member of $D(s)$.
\item $\delta$ is a \define{transition function} that assigns to every $s\in\set{S}$ and every $\sigma \in D(s)$ a state $\delta(s,\sigma) \in \set{S}$ that results from $s$ if every agent $a \in \allag$ plays move $\sigma_a$.
\end{itemize}
\end{definition}

In the following, let $\frames{F} = \Frames{\allag,\set{S},s_0,d,\delta}$ be a CGF with $s,s' \in \set{S}$. We say that $s'$ is a \define{successor} of $s$ (an $s$-successor) if $s'=\delta(s,\sigma)$, for some $\sigma\in D(s)$. If $\kappa$ is a tuple, then $\kappa_n$ (or $\kappa(n)$) denotes the $n$-th element of $\kappa$. Let $|\allag| =k$ and let $\set{A}\subseteq\allag$ be a coalition. An \define{$\set{A}$-move} $\sigma_{\scriptset{A}}$ at $s\in\set{S}$ is a $k$-tuple such that $\sigma_{\scriptset{A}}(a) \in D(a,s)$ for every $a \in \set{A}$ and $\sigma_{\scriptset{A}}(a')=*$ (i.e.\ an arbitrary move) for every $a' \not\in \set{A}$. We denote by $D(\set{A},s)$ the set of all $\set{A}$-moves at state $s$. A move vector $\sigma$ \define{extends} an $\set{A}$-move vector $\sigma_{\scriptset{A}}$, denoted by $\sigma_{\scriptset{A}} \sqsubseteq \sigma$ or $\sigma \sqsupseteq\sigma_{\scriptset{A}}$, if $\sigma(a)=\sigma_{\scriptset{A}}(a)$ for every $a \in \set{A}$. Let $\sigma_{\scriptset{A}} \in D(\set{A},s)$ be an $\set{A}$-move. The \define{outcome} of $\sigma_{\scriptset{A}}$ at $s$, denoted by $out(s,\sigma_{\scriptset{A}})$, is the set of all states $s' \in \set{S}$ for which there exists a move vector $\sigma \in D(s)$ such that $\sigma_{\scriptset{A}} \sqsubseteq \sigma$ and $\delta(s,\sigma)=s'$.

\begin{definition}\label{def:concurrent game model} A \define{Concurrent Game Model} (CGM) is a tuple $\model{M}=\Model{\frames{F},\Prop,\pi}$, where $\frames{F}=\Frames{\allag,\set{S},s_0,d,\delta}$ is a CGF; $\Prop$ is the set of propositional symbols; and $\pi:\set{S}\longrightarrow 2^{\Prop}$ is a valuation function.
\end{definition}

\begin{definition}Let $\model{M} = \Model{\allag,\set{S},s_0,d,\delta,\Prop,\pi}$ be a CGM with $s \in \set{S}$. The satisfaction relation, denoted by $\models$, is inductively defined as follows.

\begin{itemize}[noitemsep]
\item $\modelw{M}{s} \models \constant{true}$;
\item $\modelw{M}{s} \models p$ iff $p \in \pi(s)$, for all $p \in \Prop$;
\item $\modelw{M}{s} \models \neg \varphi$ iff $\modelw{M}{s} \not\models \varphi$;
\item $\modelw{M}{s} \models \varphi \land \psi$ iff $\modelw{M}{s} \models \varphi$ and $\modelw{M}{s} \models \psi$;
\item $\modelw{M}{s} \models \coop{A}\varphi$ iff there exists a $\set{A}$-move $\sigma_{\scriptset{A}} \in D(\set{A},s)$ s.t.\ $\modelw{M}{s'} \models \varphi$ for all $s'\in out(s,\sigma_{\scriptset{A}})$;
\item $\modelw{M}{s} \models \dualcoop{A}\varphi$ iff for all $\set{A}$-moves $\sigma_{\scriptset{A}} \in D(\set{A},s)$ exists $s'\in out(s,\sigma_{\scriptset{A}})$ s.t.\ $\modelw{M}{s'} \models \varphi$.
\end{itemize}
\end{definition}

\noindent Semantics of $\constant{false}$, disjunctions, and implications are given in the usual way. Given a model $\model{M}$, a state $s$ in $\model{M}$, and a formula $\varphi$, if $\modelw{M}{s} \models \varphi$, $s \in \set{S}$, we say that $\varphi$ is \define{satisfied at the state $s$ in \model{M}}.

In this work, we consider \emph{tight satisfiability}, i.e. the evaluation of a formula $\varphi$ depends only on the agents occurring in $\varphi$ \cite{WLWW2006}. We denote by $\allag_\varphi$, where $\allag_\varphi \subseteq \allag$, the set of agents occurring in a well-formed formula $\varphi$. If $\Phi$ is a set of well-formed formulae, $\allag_\Phi \subseteq \allag$ denotes $\bigcup_{\varphi \in \Phi} \allag_\varphi$. Let $\varphi \in \WFF{CL}$ and $\model{M} = \Model{\allag_\varphi,\set{S},s_0,d,\delta,\Prop,\pi}$ be a CGM. Formulae are interpreted with respect to the distinguished world $s_0$. Thus, a formula $\varphi$ is said to be \define{satisfiable in \model{M}}, denoted by $\model{M}\models \varphi$, if $\modelw{M}{s_0} \models \varphi$; it is said to be \define{satisfiable} if there is a model $\model{M}$ such that $\modelw{M}{s_0} \models \varphi$; and it is said to be \define{valid} if for all models $\model{M}$ we have $\modelw{M}{s_0} \models \varphi$. A finite set $\Gamma \subset \WFF{CL}$ is \define{satisfiable in a state $s$ in $\model{M}$}, denoted by $\modelw{M}{s}\models\Gamma$, if for all $\gamma_i \in \Gamma$, $0\leq i \leq n$, $n \in \Nat$, $\modelw{M}{s}\models \gamma_i$; $\Gamma$ is \define{satisfiable in a model $\model{M}$}, $\model{M}\models\Gamma$, if $\modelw{M}{s_0}\models\Gamma$; and $\Gamma$ is \define{satisfiable}, if there is a model $\model{M}$ such that $\model{M}\models\Gamma$.

\section{Resolution Calculus}\label{sec:calculus}

The resolution calculus for \system{CL}{}{}, \calculus{CL}, operates on sets of clauses. A formula in \system{CL}{}{} is firstly converted into a coalition problem, which is then transformed into a coalition problem in \emph{Divided Separated Normal Form for Coalition Logic}, \snf{CL}. 

\begin{definition} A \define{coalition problem} is a tuple $\Cp{\set{I},\set{U}, \set{N}}$, where $\set{I}$, the set of initial formulae, is a finite set of propositional formulae; $\set{U}$, the set of global formulae, is a finite set of formulae in \WFF{CL}; and $\set{N}$, the set of coalition formulae, is a finite set of coalition formulae, i.e.\ those formulae in which a coalition modality occurs.
\end{definition}

The semantics of coalition problems assumes that initial formulae hold at the initial state; and that global and coalition formulae hold at every state of a model. 

\begin{definition} Given a coalition problem $\cp{C}=\Cp{ \set{I},\set{U}, \set{N}}$, we denote by $\allag_{\scriptcp{C}}$ the set of agents $\allag_{\scriptset{U} \cup \scriptset{N}}$. If $\cp{C}=\Cp{ \set{I},\set{U}, \set{N}}$ is a coalition problem and $\model{M} = \Model{\allag_{\scriptcp{C}},\set{S},s_0,d,\delta,\Prop,\pi}$ is a CGM, then $\model{M}\models \cp{C}$ if, and only if, $\modelw{M}{s_0} \models \set{I}$ and $ \modelw{M}{s} \models \set{U}\cup\set{N}$, for all $s \in \set{S}$. We say that $\cp{C}=\Cp{ \set{I},\set{U}, \set{N}}$ is \define{satisfiable}, if there is a model $\model{M}$ such that $\model{M}\models \cp{C}$. 
\end{definition}

In order to apply the resolution method, we further require that formulae within each of those sets are in \emph{clausal form}: \define{initial clauses} and \define{global clauses} are of the form $\bigvee_{j=1}^n l_j$; \define{positive coalition clauses} are of the form $\bigwedge_{i=1}^m l'_i  \then \coop{A} \bigvee_{j=1}^n l_j$; and \define{negative coalition clauses} are of the form $\bigwedge_{i=1}^m l'_i \then \dualcoop{A} \bigvee_{j=1}^n l_j$; where $m,n \geq 0$ and $l'_i, l_j$, for all $1 \leq i \leq m$,  $1 \leq j \leq n$, are literals or constants. We assume that clauses are kept in the simplest form by means of usual Boolean simplification rules. Tautologies are removed from the set of clauses as they cannot contribute to finding a contradiction. A \define{coalition problem in \snf{CL}} is a coalition problem $\Cp{\set{I},\set{U}, \set{N}}$ such that \set{I} is a set of initial clauses, \set{U} is a set of global clauses, and \set{N} is a set of positive and negative coalition clauses.

The transformation of a coalition logic formula into a coalition problem in \snf{CL} is analogous to the approach taken in \cite{DFK06}. The transformation of a formula into a \define{coalition problem in \snf{CL}}, which is given in \cite{NZDH-TR-2013,NZDH-JLC-2014}, reduces the number of operators and separates the contexts to which the resolution inference rules are applied, but may add new propositional symbols.

The set of inference rules for \calculus{CL} are given as follows. Let $\Cp{ \set{I},\set{U}, \set{N}}$ be a coalition problem in \snf{CL}; $C,C'$ be conjunctions of literals; $D,D'$ be disjunctions of literals; $l, l_i$ be literals; and $\set{A},\set{B}\subseteq\allag$ be coalitions (where $\allag$ is the set of all agents). The first rule, \rulename{IRES1}, is classical resolution applied to clauses which are true at the initial state. The next inference rule, \rulename{GRES1}, performs resolution on clauses which are true in all states.

\begin{center}
$
\begin{array}{lll}
\mbox{\rulename{IRES1}} & D \lor l & \in \set{I}\\
                        & D' \lor \neg l & \in \set{I}\cup\set{U}\\ \cline{2-2}
                        & D \lor D' & \in \set{I}\\
\end{array}
$\hspace{5ex}
$
\begin{array}{lll}
\mbox{\rulename{GRES1}} & D \lor l & \in \set{U}\\
                        & D' \lor \neg l & \in\set{U}\\ \cline{2-2}
                        & D \lor D' & \in \set{U}\\
\end{array}
$
\end{center}

\noindent Soundness of \rulename{IRES1} and \rulename{GRES1} follow from the semantics of coalition problems and the soundness result for classical propositional resolution \cite{Rob65}. The following rules perform resolution on positive and negative coalition clauses.

\begin{center}
$
\begin{array}{ll}
\begin{array}{lrcll}

\mbox{\rulename{CRES1}}       & C          & \then & \coop{A}(D \lor l) & \in \set{N}\\
\set{A}\cap\set{B}=\emptyset  & C'         & \then & \coop{B}(D'\lor \neg l) & \in \set{N} \\ \cline{2-4}
                              & C \land C' & \then & \coopl{\set{A}\cup\set{B}}(D \lor D') & \in \set{N}
\end{array}
&\begin{array}{lrcll}
\mbox{\rulename{CRES2}} &   &       & D \lor l  & \in \set{U} \\
                        & C & \then & \coop{A}(D'\lor \neg l)  & \in \set{N} \\ \cline{2-4}
                        & C & \then & \coop{A}(D \lor D') & \in \set{N}
\end{array}
\\
\\

\begin{array}{lrcll}
\mbox{\rulename{CRES3}}       & C & \then & \coop{A}(D \lor l) & \in \set{N}\\
\set{A}\subseteq\set{B}       & C' & \then & \dualcoop{B}(D'\lor \neg l) & \in \set{N}\\ \cline{2-4}
                              & C \land C' & \then & \dualcoopl{\set{B}\setminus\set{A}}(D \lor D')& \in \set{N}
\end{array}
&
\begin{array}{lrcll}
\mbox{\rulename{CRES4}} &   &       & D \lor l & \in \set{U}\\
                        & C & \then & \dualcoop{A}(D'\lor \neg l) & \in \set{N}\\ \cline{2-4}
                        & C & \then & \dualcoop{A}(D \lor D')& \in \set{N}
\end{array}
\end{array}
$
\end{center}

\noindent Soundness of the inference rules \rulename{CRES1-4} follow from the axiomatisation of \system{CL}{}{}, given in Section~\ref{sec:language}. We give sketches of the proofs here. Let $\model{M}$ be a CGM and $s \in \model{M}$ a state. Recall that coalition clauses are satisfied at any state in $\model{M}$. For \rulename{CRES1}, if $\modelw{M}{s}\models C \land C'$, by the semantics of conjunction and implication, we have that $\modelw{M}{s}\models C \land C'\then  \coop{A}(D \lor l) \land \coop{B}(D' \lor \neg l)$. By axiom \axiom{S}, we have that $ \coop{A}(D \lor l) \land \coop{B}(D' \lor \neg l)$ implies $\coopl{\set{A}\cup\set{B}}((D \lor l) \land (D' \lor \neg l))$. Therefore, $\modelw{M}{s}\models C \land C'\then \coopl{\set{A}\cup\set{B}}((D \lor l) \land (D' \lor \neg l))$. By classical resolution applied within the successor states, we obtain that $\modelw{M}{s}\models C \land C'\then \coopl{\set{A}\cup\set{B}}(D \lor D')$. For \rulename{CRES3}, by Lemma~\ref{lemma:axiomatisation}, we have that $\coop{A}(D \lor l) \land \dualcoop{B} (D' \lor \neg l)\then \dualcoopl{\set{B}\setminus\set{A}}((D \lor l) \land (D' \lor \neg l))$, with $\set{A}\subseteq\set{B}$, is valid. If $\modelw{M}{s}\models C \land C'$, by the semantics of implication, we have that $\modelw{M}{s} \models C \land C'\then \dualcoopl{\set{B}\setminus\set{A}}((D \lor l) \land (D' \lor \neg l))$. Applying classical resolution within the successor states, we obtain that $\modelw{M}{s} \models C \land C'\then \dualcoopl{\set{B}\setminus\set{A}}(D \lor D')$. Soundness of the inference rules \rulename{CRES2} and \rulename{CRES4} follow from the above and the semantics of coalition problems: as $D \lor l$ in $\set{U}$ is satisfied at all states, we have that $\constant{true} \then\coopl{\emptyset}(D \lor l)$ is also satisfied at all states.

The next two inference rules are justified by the axioms \axiom{$\bot$} and \axiom{$\top$}, given by $\neg\coop{A}\constant{false}$ and $\coop{A}\constant{true}$, respectively, which imply that the consequent in both rewriting rules cannot be satisfied. 

\begin{center}
$
\begin{array}{lcll}
\mbox{\rulename{RW1}} & \bigwedge_{i=1}^n l_i \then \coop{A}\constant{false} & \in \set{N}\\ \cline{2-2}
           &  \bigvee_{i=1}^n \neg l_i & \in \set{U} &
\end{array}
$\hspace{5ex}
$
\begin{array}{lcll}
\mbox{\rulename{RW2}} & \bigwedge_{i=1}^n l_i \then \dualcoop{A}\constant{false} & \in \set{N}\\ \cline{2-2}
           & \bigvee_{i=1}^n \neg l_i & \in \set{U} &
\end{array}
$
\end{center}

As sketched above, the resolution-based calculus for Coalition Logic is sound.

\begin{theorem}[Soundness]\label{theo:inference-rules}%
Let $\cp{C}$ be a coalition problem in \snf{CL}. Let $\cp{C}'$ be the coalition problem in \snf{CL} obtained from $\cp{C}$ by applying any of the inference rules \rulename{IRES1}, \rulename{GRES1}, \rulename{CRES1-4} or \rulename{RW1-2} to $\cp{C}$. If $\cp{C}$ is satisfiable, then $\cp{C}'$ is satisfiable.
\end{theorem}

A \define{derivation} from a coalition problem in \snf{CL} $\cp{C}=\Cp{ \set{I},\set{U},\set{N}}$ by $\calculus{CL}$ is a sequence $\cp{C}_0, \cp{C}_1,\\ \cp{C}_2, \ldots$ of problems such that $\cp{C}_0 = \cp{C}$, $\cp{C}_i=\Cp{ \set{I}_i,\set{U}_i,\set{N}_i}$, and $\cp{C}_{i+1}$ is either $\Cp{ \set{I}_i \cup \Set{D},\set{U}_i,\set{N}_i }$, where $D$ is the conclusion of\/ \rulename{IRES1}; $\Cp{ \set{I}_i ,\set{U}_i\cup \Set{D},\set{N}_i }$, where $D$ is the conclusion of\/ \rulename{GRES1}, \rulename{RW1}, or\/ \rulename{RW2}; or $\Cp{ \set{I}_i ,\set{U}_i,\set{N}_i\cup \Set{D}}$, where $D$ is the conclusion of\/ \rulename{CRES1}, \rulename{CRES2}, \rulename{CRES3}, or\/ \rulename{CRES4}; and $D$ is not a tautology.

A \define{refutation} for a coalition problem in \snf{CL} $\cp{C}=\Cp{ \set{I},\set{U},\set{N}}$ (by\/ $\calculus{CL}$\/) is a derivation from $\cp{C}$ such that for some $i\geq 0$, $\cp{C}_i=\Cp{\set{I}_i,\set{U}_i,\set{N}_i}$ contains a contradiction, where a contradiction is given by either $\constant{false}\in \set{I}_i$ or $\constant{false}\in \set{U}_i$. A derivation \emph{terminates} if, and only if, either a contradiction is derived or no new clauses can be derived by further application of resolution rules of $\calculus{CL}$.

The completeness proof for \calculus{CL} is based on the tableau construction given in \cite{GS2009}. Given an unsatisfiable coalition problem in \snf{CL} $\cp{C}$, an initial tableau is obtained by this construction which is then reduced to an empty tableau via a sequence of deletion steps. We show that each deletion step corresponds to an application of the resolution inference rules to (sub)sets of clauses in $\cp{C}$ or clauses previously derived from $\cp{C}$. The derivation constructed in this way is shown to be a refutation of $\cp{C}$.


\begin{theorem}[Completeness]\label{theorem-completeness}
Let $\cp{C}=\Cp{\set{I},\set{U},\set{N}}$ be an unsatisfiable coalition problem in \snf{CL}. Then there is a refutation for $\cp{C}$ using the inference rules \rulename{IRES1}, \rulename{GRES1}, \rulename{CRES1-4}, and \rulename{RW1-2}.
\end{theorem}

The proof that every derivation terminates is trivial and based on the fact that we have a finite number of clauses that can be expressed. As the number of propositional symbols after translation into the normal form is finite and the inference rules do not introduce new propositional symbols, we have that the number of possible literals occurring in clauses is finite and the number of conjunctions (resp.\ disjunctions) on the left-hand side (resp.\ right-hand side) of clauses is finite (modulo simplification). As the number of agents is finite, the number of coalition modalities that can be introduced by inference rules is also finite. Thus, only a finite number of clauses can be expressed (modulo simplification), so at some point either we derive a contradiction or no new clauses can be generated.

\begin{theorem}\label{theo:termination}%
Let $\cp{C}=\Cp{\set{I},\set{U},\set{N}}$ be a coalition problem in \snf{CL}. Then any derivation from $\cp{C}$ by \calculus{CL} terminates.
\end{theorem}

Full proofs for soundness, completeness, termination, and complexity of the resolution-based method for \system{CL}{}{} are given in \cite{NZDH-TR-2013,NZDH-JLC-2014}.

\begin{example}\label{example} We show a simple example, adapted from \cite{Herzig-ESSLLI-2007}, of the application of \calculus{CL} to a problem involving the cooperation of agents. There are two agents ($1$ and $2$) and two toggle switches. For each agent $\agent=1,2$, there are two possible actions: $\coopl{\agent}tog_\agent \land \coopl{\agent} \neg tog_\agent$, where $tog_\agent$ denotes that the agent $\agent$ can toggle the switch $\agent$ (clauses 3, 9--13). The light is initially off, i.e.\ we have that $t_0 \then \neg l$ (clauses 1 and 2). If the light is off and the switch is toggled, then at the next moment the light is on: $tog_\agent \land \neg l \then \coopl{\agent} l$ (clauses 5 and 6). Similarly, if the light is on and the agent toggles the switch, then at the next moment the light is off: $tog_\agent \land l \then \coopl{\agent} \neg l$ (clauses 7 and 8). We prove that the agents can cooperate to turn on the light, that is, we introduce the clauses 4 and 14, which corresponds to the negation of $\coopl{1,2} l$.\vspace{2ex}

\hspace{-5.5ex}\begin{raggedleft}
\begin{tabular}{rrcll}
1. &  & & $t_0$ & $\jusinit{I}$ \\
2. &  & & $\neg t_0 \lor \neg l$ & $\jusinit{U}$ \\
3. &  & & $\neg t_0 \lor t_1$ & $\jusinit{U}$ \\
4. &  & & $\neg t_1 \lor t_4$ & $\jusinit{U}$ \\
5. & $tog_1 \land \neg l$ & $\then$ & $\coopl{1}l$ & $\jusinit{N}$ \\
6. & $tog_2 \land \neg l$ & $\then$ & $\coopl{2}l$ & $\jusinit{N}$ \\
7. & $tog_1 \land  l$ & $\then$ & $\coopl{1}\neg l$ & $\jusinit{N}$ \\
8. & $tog_2 \land  l$ & $\then$ & $\coopl{2}\neg l$ & $\jusinit{N}$ \\
9. & $t_1$ & $\then$ & $\coopl{1}tog_1$ & $\jusinit{N}$ \\
10. & $t_1$ & $\then$ & $\coopl{2}tog_2$ & $\jusinit{N}$ \\
11. & $t_1$ & $\then$ & $\coopl{1}\neg tog_1$ & $\jusinit{N}$ \\
12. & $t_1$ & $\then$ & $\coopl{2}\neg tog_2$ & $\jusinit{N}$ \\
13. & $t_1$ & $\then$ & $\coopl{\emptyset}t_1$ & $\jusinit{N}$ \\
\end{tabular}
\hspace{-1ex}\begin{tabular}{rrcll}
14. & $t_4$ & $\then$ & $\coopl{\emptyset}\neg l$ & $\jusinit{N}$\\
15. &  & & $\neg t_0 \lor t_4$ & $\jus{U}{gres1}{3, 4}$ \\
16. & $t_4 \land tog_1 \land \neg l$ & $\then$ & $\coopl{1}\constant{false}$ & $\jus{N}{cres1}{5, 14}$ \\
17. & $t_1$ & $\then$ & $\coopl{\emptyset}t_4$ & $\jus{N}{cres2}{13, 4}$ \\
18. &  & & $l \lor \neg t_4 \lor \neg tog_1$ & $\jus{U}{rw1}{16}$ \\
19. & $t_1$ & $\then$ & $\coopl{\emptyset}l \lor \neg tog_1$ & $\jus{N}{cres2}{17, 18}$ \\
20. & $t_1$ & $\then$ & $\coopl{1}l$ & $\jus{N}{cres1}{19, 9}$ \\
21. & $t_1 \land t_4$ & $\then$ & $\coopl{1}\constant{false}$ & $\jus{N}{cres1}{20, 14}$ \\
22. &  & & $\neg t_1 \lor \neg t_4$ & $\jus{U}{rw1}{21}$ \\
23. &  & & $\neg t_0 \lor \neg t_1$ & $\jus{U}{gres1}{22, 15}$ \\
24. &  & & $\neg t_0$ & $\jus{U}{gres1}{23, 3}$ \\
25. &  & & $\constant{false}$ & $\jus{I}{ires1}{1, 24}$ \\
\\
\end{tabular}
\end{raggedleft}
\end{example}

\section{\clprover}\label{sec:implementation}

\clprover is a prototype implementation of the resolution-based method given in \cite{NZDH-JLC-2014}. The prover is written in SWI-Prolog (Multi-threaded, 64 bits, Version 6.0.2) and the compiled binaries for Linux x86\_64 together with instructions for usage and example files are available at \url{http://www.cic.unb.br/docentes/nalon/#software}. 

The prover recurs over the set of clauses using breadth-first search for a proof. The resolution inference rules for \system{CL}{}{} are in fact variations of the propositional resolution rule. Before presenting the general form of the inference rules, we explain the data structures that are employed by the prover. A \emph{clause core} is implemented as a list with three elements, all of which are lists: a list of literals on the left-hand side of a clause, a list of agents, and a list of literals on the right-hand side of a clause. The only operator allowed within the lists of literals is the negation operator, \verb+neg+. Clauses are then given as Prolog lists, with four elements. The first element is the clause number, the second is the clause core, the third is the justification (`given', if the clause is an input clause; or a list containing the numbers of the clauses from which it was derived, together with the literal being resolved, and the inference rule applied), and the fourth is an indication to which set within a coalition problem the clause belongs (`i' for initial, `u' for global, `p' for positive, and `n' for negative). Thus, for instance, the clauses 1, 3, and 20 from Example~\ref{example} are represented as \verb+[1,[[],[],[t0]],[given],i]+, \verb+[3,[[],[],[neg t0, t1]],[given],u]+, and \verb+[20,[[t1],[1],[l]],[9,19,tog1,cres1],n]+, respectively. 

Given this representation, the propositional resolution inference rule is modified in such a way that a clause \verb+[_,[LHS1],[AG1],[RHS1],_,S1]+ is resolved with \verb+[_,[LHS2],[AG2],[RHS2],_,S2]+, if such clauses meet the side conditions given by the inference rules presented in Section~\ref{sec:calculus}. For instance, the rule \rulename{CRES1} is applied if both \verb+S1+ and \verb+S2+ are equal to 'p' and if the intersection between \verb+LHS1+ and \verb+LHS2+ is empty. The prover then recurs over the set of initial, global and coalition clauses using the following procedure (where S is a saturated set of clauses and N is a non-saturated set of clauses):

\begin{center}
{
\begin{minipage}{.4\textwidth}
{\bf procedure} resolution(S, N)

{\bf while} (N $\neq \emptyset$ and $\constant{false} \not\in$  N)

{\bf do} \hspace{.1ex} Given $\leftarrow$ choose(N);

\hspace{3ex} N $\leftarrow$ N $\setminus$ \{Given\};

\hspace{3ex} S $\leftarrow$ S $\cup$ \{Given\};

\hspace{3ex} New $\leftarrow$ rewrite(res(Given,S));

\hspace{3ex}\verb+/* Forward Subsumption */+

\hspace{3ex} N $\leftarrow$ sub(sub(New,S),N);

{\bf end-while}

{\bf if} $\constant{false} \in$ N {\bf then } S $\leftarrow$ S $\cup$ \{\constant{false}\};

{\bf return} S;
 \end{minipage}
}
\end{center}

\noindent where choose(N) randomly picks a clause in N; res(C, N) is the set of all non-tautological resolvents in simplified form derivable between a clause C and a set of clauses N by one of the inference rules; rewrite(N) is the union of N and the set of clauses derived by the rewriting rules; and sub(M,N) is the set of clauses in M not subsumed by a clause in N. Forward subsumption is implemented for both the propositional and modal portions of the language. For the propositional part, a clause $D$ in $\set{I}$ (resp.\ $\set{U}$) is subsumed by a clause $D'$ in $\set{I}\cup\set{U}$ (resp.\ $\set{U}$) if $D' \subseteq D$. A positive coalition clause $C \then \coop{A} D$ is subsumed by another positive coalition clause $C' \then \coop{A'} D'$, if $C' \subseteq C$, $\set{A}' \subseteq \set{A}$, and $D' \subseteq D$. A negative coalition clause $C \then \dualcoop{A} D$ is subsumed by another negative coalition clause $C' \then \dualcoop{A'} D'$, if $C' \subseteq C$, $\set{A} \subseteq \set{A}'$, and $D' \subseteq D$. Some other forms of subsumption have not been implemented in the current version of the prover, as, for instance, coalition clauses which are subsumed by global clauses.
  
The current version of \clprover is a prototype. The prover implements subsumption, but it does not implement any of the usual performance improving techniques for resolution-based methods. For example, the function choose(N) does not use any heuristic to determine what given clause to pick. Further refinements of the resolution calculus, for example, ordered resolution or the the use of a set of support strategy would also greatly improve the performance of the prover. \clprover, however, performs well when compared with both versions of another tool, namely, TATL, a tableau-based prover for \system{ATL}{}{} \cite{AD2013:TATL}. In the following, TATL-A refers to the April version of the TATL prover, available at at \url{https://www.ibisc.univ-evry.fr/~adavid/bin/tatl.tar.gz}; and TATL-N refers to the November version, available at \url{http://atila.ibisc.univ-evry.fr/tableau_ATL/bin/tatl.tar.gz}.

A benchmark, consisting of six sets of randomly generated \system{CL}{}{} formulae, was designed to compare the performance of both provers. The formulae in the benchmark are characterised by five parameters: (1) the number of propositional symbols $N$; (2) the number of agents $A$; (3) the number of conjuncts $L$; (4) the modal degree $D$; and (5) the probability $P$. Based on a given choice of parameters random formulae in conjunctive normal form (CNF) are defined inductively as follows. A \emph{random (coalition) atom} of degree $0$ is a propositional variable randomly chosen from the set of $N$ propositional symbols. A \emph{random coalition atom} of degree $D$, $D > 0$, is with probability $P$: (a) an expression of the form $\coop{A}\varphi$, where $\coop{A}$ is a coalition modality with a set of agents $\set{A}$ randomly chosen from $\set{P}^{\{1,\ldots,A\}}$ and $\varphi$ is a random coalition CNF clause of modal degree $D-1$ (defined below), or (b) a random atom of degree $0$, otherwise. A \emph{random coalition literal} (of degree $D$) is with probability $0.5$ a random coalition atom (of degree $D$) or its negation, otherwise. A \emph{random coalition CNF clause} (of degree $D$) is a disjunction of
three random coalition literals (of degree $D$). A \emph{random coalition CNF formula} (of degree $D$) is a conjunction of $L$ random coalition CNF clauses (of degree $D$).

\begin{figure}[t]
\centering
\includegraphics[width=.9\textwidth]{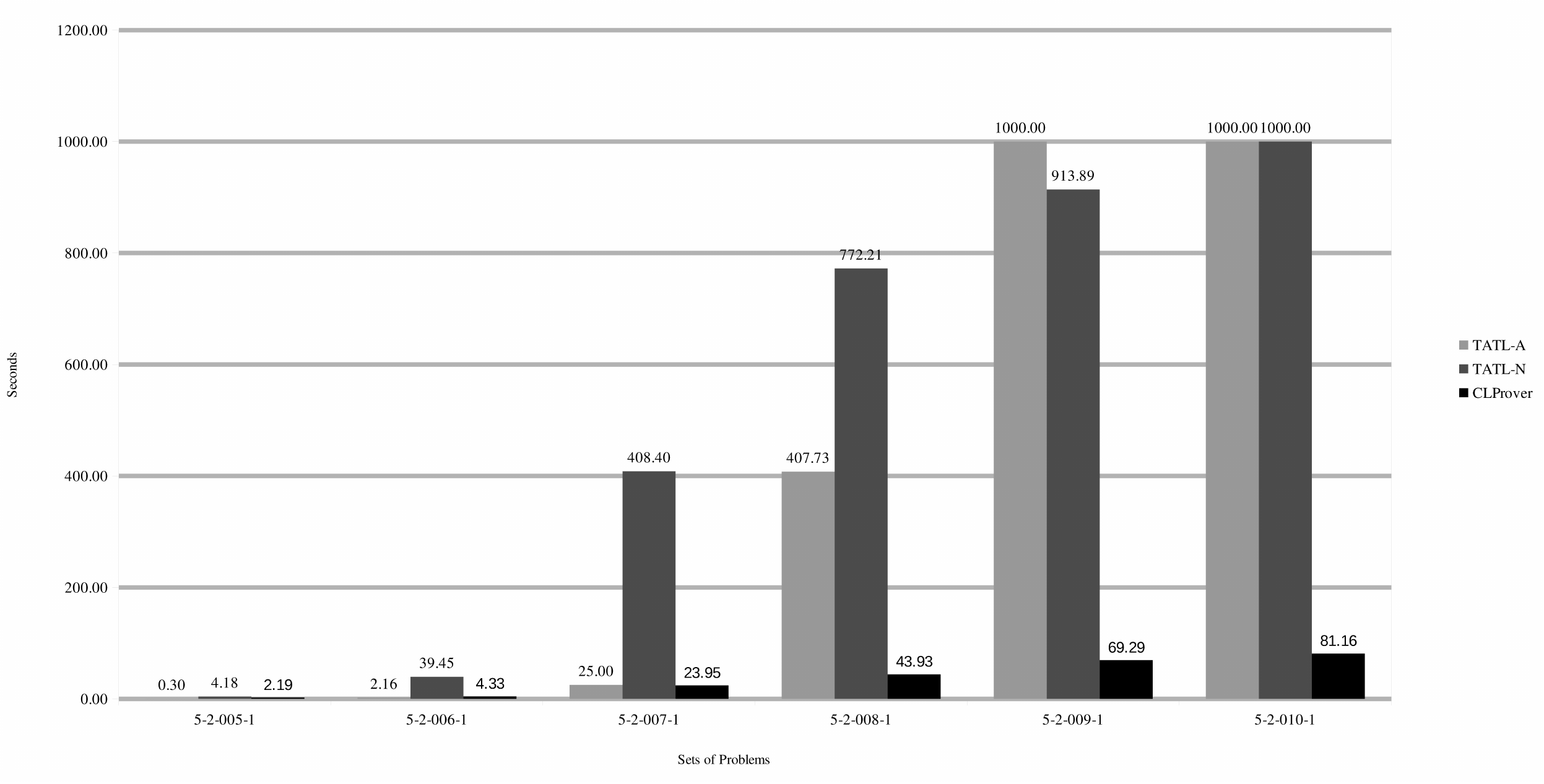}
\caption{Performance Comparison between \protect\clprover, TATL-A, and TATL-N.}
\label{fig1}
\end{figure}

The six sets of problems used to compare \clprover and TATL were generated using $N=5$, $A=2$, $L \in\{5,\ldots, 10\}$, $D=1$, and $P=1$. The experiment was run on an i7-3537U CPU at 3.00GHz, 8GB RAM, under Linux kernel 3.11.10-100. The average run-time for each set of problems is shown in Figure~\ref{fig1}. The provers were given a timeout of 1000 seconds. \clprover has solved all problems in all sets within the given time. TATL-A has failed to solve any problems in the sets 5-2-009-1 and 5-2-010-1. TATL-N has solved all problems in the sets 5-2-005-1 and 5-2-006-1; nine problems in 5-2-007-1; three in 5-2-008-1; four in 5-2-009-1; and none in 5-2-010-1. For the calculation of the average times, whenever the prover has timed out, we have set the corresponding time to 1000 seconds.

\section{Conclusion}\label{sec:conclusion}

The resolution-based calculus for the Coalition Logic \system{CL}{}{} is applied to a coalition problem in \snf{CL}, which separates the dimensions to which the resolution rules are applied. The transformation into the normal form is satisfiability preserving and polynomially bounded by the size of the original formula. Soundness of the method follows from the axiomatisation of \system{CL}{}{}. Completeness is proved with respect to the tableau procedure given in \cite{GS2009}: if a tableau for a coalition problem is closed, there is a refutation based on the calculus given here. Termination is ensured by the fact that number of propositional symbols and agents is finite, so there are only a finite number of clauses that can be generated.

The decision procedure based on \calculus{CL} is in EXPTIME, as shown in \cite{NZDH-JLC-2014}. This is optimal, as the satisfiability problem for coalition problems in \snf{CL} is EXPTIME-hard, thus more expressive than the language of \system{CL}{}{}. This result follows from \cite[Lemma 4.10, page 785]{WLWW2006} and the fact that an extension of \system{CL}{}{} with positive occurrences of \system{ATL}{}{}'s $\excoal{\emptyset}\always$ operator can be translated into \snf{CL}. It also follows that \snf{CL} is more expressive than \system{CL}{}{}.


\clprover is the first (prototype) implementation of \calculus{CL}. The experiments we have performed suggest that it is a viable tool for reasoning about Coalition Logic. Future work includes further improvements to \clprover. We also intend to extend our calculus to the full language of \system{ATL}{}{}.


\bibliographystyle{eptcs}
\bibliography{atl}
\end{document}

%% file: symbols.tex
\newcommand{\comment}[1]{}
\newcommand{\define}[1]{{\bfseries #1}}

\newcommand{\agent}{\ensuremath{a}}

\makeatletter
\def\provideenvironment{\@star@or@long\provide@environment}
\def\provide@environment#1{%
        \@ifundefined{#1}%
                {\def\reserved@a{\newenvironment{#1}}}%
                {\def\reserved@a{\renewenvironment{#1}}}%
        \reserved@a
}
\def\dummy@environ{}
\makeatother

\provideenvironment{proof}[1]{\hbox{\bf Proof of Theorem~#1}\noindent }{$\Box$\vspace{1ex}}
\provideenvironment{lproof}{\hbox{\bf Proof.}\noindent }{}

\newcounter{definition}
\newenvironment{definition}{\refstepcounter{definition}{\bf Def. \thedefinition} }{}

\newtheorem{example}{Example}

\ifx\fmtname\@psfmtname \else \def\cmsy@{2}\fi 
\def\sometime{\mathord{\hbox{\large$\mathchar"0\cmsy@7D$}}}

\newcommand{\then}{\Rightarrow}

\newcommand{\ifonlyif}{\Leftrightarrow}

\newcommand{\constant}[1]{\mbox{\rm\bf #1}}
\newcommand{\system}[3]{\raisebox{.2ex}[1.2ex]{\raisebox{-.2ex}{{\sf #1}}{$_{#2}^{#3}$}}}

\newcommand{\set}[1]{\mbox{\ensuremath{\mathcal{#1}}}}
\newcommand{\scriptset}[1]{\mbox{\scriptsize\ensuremath{\mathcal{#1}}}}
\newcommand{\Set}[1]{\ensuremath{\{#1\}}}

\newcommand{\frames}[1]{\ensuremath{\mathcal{#1}}}
\newcommand{\Frames}[1]{\ensuremath{(#1)}}

\newcommand{\Tuple}[1]{\ensuremath{(#1)}}

\newcommand{\model}[1]{\ensuremath{\set{#1}}}
\newcommand{\modelw}[2]{\ensuremath{\langle \model{#1}, #2 \rangle}}
\newcommand{\Model}[1]{\Tuple{#1}}

\newcommand{\scriptcp}[1]{\ensuremath{\scriptset{#1}}}
\newcommand{\cp}[1]{\ensuremath{\set{#1}}}
\newcommand{\Cp}[1]{\ensuremath{\Tuple{#1}}}

\newcommand{\WFF}[1]{{\sf WFF}{\mbox{$_{\mbox{\small\sf #1}}$}}}

\newcommand{\Prop}{\ensuremath{\Pi}} 

\newcommand{\Nat}{\mbox{$\mathbb N$}}

\newcommand{\always}{\raisebox{-.2ex}{
			   \mbox{\unitlength=0.9ex
			   \begin{picture}(2,2)
			   \linethickness{0.06ex}
			   \put(0,0){\line(1,0){2}}
			   \put(0,2){\line(1,0){2}}
			   \put(0,0){\line(0,1){2}}
			   \put(2,0){\line(0,1){2}}
			   \end{picture}}}
		      \,}

\newcommand{\Next}{\!\raisebox{-.2ex}{ 
			\mbox{\unitlength=0.9ex
			\begin{picture}(2,2)
			\linethickness{0.06ex}
			\put(1,1){\circle{2}} 
			\end{picture}}}	      
			\,}

\newcommand{\coop}[1]{\ensuremath{[\set{#1}]}}
\newcommand{\coopl}[1]{\ensuremath{[ #1 ]}}
\newcommand{\dualcoop}[1]{\ensuremath{{\langle}\set{#1}{\rangle}}} 
\newcommand{\dualcoopl}[1]{\ensuremath{{\langle}#1{\rangle}}}

\newcommand{\excoa}[1]{\ensuremath{\langle\!\langle{\set{#1}}\rangle\!\rangle}}     

\newcommand{\excoal}[1]{\ensuremath{\langle\!\langle{#1}\rangle\!\rangle}}     

\newcommand{\allag}{\ensuremath{\Sigma}}

\newcommand{\snf}[1]{{\sf DSNF}\raisebox{-.5ex}{{\scriptsize{\sf #1}}}}
\newcommand{\calculus}[1]{{\sf RES}\raisebox{-.5ex}{{\scriptsize{\sf #1}}}}
\newcommand{\rulename}[1]{{\sf\bf #1}}
\newcommand{\axiom}[1]{{\sf\bf #1}}
\newcommand{\jus}[3]{\mbox{$[$\set{#1}\!,\uppercase{\rulename{#2}},#3$]$}}
\newcommand{\jusinit}[1]{\mbox{$[$\set{#1}$]$}}
